\documentclass[12pt]{article}
\usepackage[dvips]{color}
\usepackage{epsfig}
\usepackage{amsmath}
\usepackage{graphicx}
\begin{document}
\title{{\bf First-order formalism for the quintom model of dark energy }}
\author{  M. R. Setare \thanks{%
E-mail: rezakord@ipm.ir}\\{Department of Science, Payame Noor University, Bijar, Iran} \\
 J. Sadeghi\thanks{%
E-mail: pouriya@ipm.ir
 }\\
{ Sciences Faculty, Department of Physics, Mazandaran University
}\\{  P .O .Box 47415-416, Babolsar, Iran}}

 \maketitle
\begin {abstract}
The present paper deals to the quintom model of dark energy. We
introduce a first-order formalism, which shows how to relate the
potential that specifies the scalar field model to Hubble parameter.
Reviewing briefly the quintom scenario of dark energy, we present a
general procedure to solve the equations of motion for quintom model
driven by a couple scalar fields with first-order differential
equations.
\end {abstract}
\newpage
\section{Introduction}
Recent observations from type Ia supernovae \cite{SN} in associated
with Large Scale Structure \cite{LSS} and Cosmic Microwave
Background anisotropies \cite{CMB} have provided main evidence for
the cosmic acceleration. The combined analysis of cosmological
observations suggests that the universe consists of about $70\%$
dark energy, $30\%$ dust matter (cold dark matter plus baryons), and
negligible radiation. Although the nature and origin of dark energy
are unknown, we still can propose some candidates to describe it.
The most obvious theoretical candidate of dark energy is the
cosmological constant $\lambda$ (or vacuum energy)
\cite{Einstein:1917,cc} which has the equation of state $w=-1$.
However, as is well known, there are two difficulties arise from the
cosmological constant scenario, namely the two famous cosmological
constant problems --- the ``fine-tuning'' problem and the ``cosmic
coincidence'' problem \cite{coincidence}. An alternative proposal
for dark energy is the dynamical dark energy scenario. The dynamical
dark energy proposal is often realized by some scalar field
mechanism which suggests that the energy form with negative pressure
is provided by a scalar field evolving down a proper potential.\\
So far, a large class of scalar-field dark energy models have been
studied, including quintessence \cite{quintessence}, K-essence
\cite{kessence}, tachyon \cite{tachyon}, phantom \cite{phantom},
ghost condensate \cite{ghost2} and quintom \cite{quintom}, and so
forth. In addition, other proposals on dark energy include
interacting dark energy models \cite{intde}, Chaplygin gas models
\cite{cg},holographic dark energy \cite{holoext}, and many others.
Recently there are many relevant studies on phantom energy
\cite{meng}. The analysis of the properties of dark energy from
recent observations mildly favor models with $w$ crossing -1 in the
near past. But, neither quintessence nor phantom can fulfill this
transition. In the quintessence model, the equation of state
$w=p/\rho$ is always in the range $-1\leq w\leq 1$ for $V(\phi)>0$.
Meanwhile for the phantom which has the opposite sign of the kinetic
term compared with the quintessence in the Lagrangian, one always
has $w\leq -1$. Neither the quintessence nor the phantom alone can
fulfill the transition from $w>-1$ to $w<-1$ and vice versa.
Although for k-essence\cite{kessence} one can have both $w\ge -1$
and $w<-1$, it has been lately considered by Ref\cite{Vikman1,
Vikman2} that it is very difficult for k-essence to get $w$ across
$-1$ during evolving. But one can show \cite{{quintom},{quint2}}
that considering the combination of quintessence and phantom in a
joint model, the transition can be fulfilled. This model, dubbed
quintom, can produce a better fit to the data than more familiar
models with $w\geq-1$. In the other term the quintom model of dark
energy represents a transition of dark energy equation of state from
$w>-1$ to $w<-1$, or vice versa, namely from $w<-1$ to $w>-1$ is
also one realization of quintom, as can be seen clearly in
\cite{ref}. \\
In this paper we focus attention on the quintom model of dark energy
by first-order formalism \cite{bez}(see also \cite{bez2}), which
shows how to relate the potential that specifies the scalar field
model to Hubble's parameter. By using \cite{bez} we can continue
this process for the two fields in quintom model, in another term we
present a general procedure to solve the equations of motion for
quintom model driven by a couple scalar fields with first-order
differential equations.
\section{The quintom model of dark energy }
The quintom model of dark energy \cite{quint2} is of new models
proposed to explain the new astrophysical data, due to transition
from $w>-1$ to $w<-1$, i.e. transition from quintessence dominated
universe to phantom dominated universe. Here we consider the
spatially flat Friedman-Robertson-Walker universe, where has
following space-time metric,
\begin{equation}\label{met}
ds^{2}=-dt^{2}+a(t)^{2}(dr^{2}+r^{2}d\Omega^{2}).
\end{equation}
Containing the normal scalar field $\sigma$ and negative kinetic
scalar field $\phi$, the action which describes the quintom model is
expressed as the following form,
\begin{equation}\label{1}
S=\int d^4x\sqrt{-g}\left(\frac{R}{4}
 -\frac{1}{2}g^{\mu \nu}\partial _\mu \phi \partial _\nu \phi
 +\frac{1}{2}g^{\mu \nu}\partial _\mu \sigma \partial _\nu \sigma
 -V(\phi ,\sigma)\right),
\end{equation}
where we have not considered the lagrangian density of matter field
and we are using $4\pi G=1$. In the spatially flat
Friedman-Robertson-Walker (FRW) universe, the effective energy
density, $\rho$, and the effective pressure, P, of the scalar fields
can be described by;
\begin{eqnarray}\label{2}
\rho=-\frac{1}{2}\dot{\phi}^2+\frac{1}{2}\dot{\sigma}^2+V(\phi,\sigma), \\
P=-\frac{1}{2}\dot{\phi}^2+\frac{1}{2}\dot{\sigma}^2-V(\phi,\sigma).
\end{eqnarray}
So, the equation of state can be written as,
\begin{equation}\label{3}
w=\frac{-\dot{\phi}^2+\dot{\sigma}^2-2V(\phi,\sigma)}
{-\dot{\phi}^2+\dot{\sigma}^2+2V(\phi,\sigma)}.
\end{equation}
From the equation of state , it is seen that for
$\dot\sigma>\dot\phi$, $w\geq -1$ and for $\dot\sigma<\dot\phi$, we
will have, $w<-1$. So, the evolution equation for two scalar fields
in FRW model will have the following form,
\begin{eqnarray}\label{5}
\ddot{\phi}+3H\dot{\phi}-\frac{dV(\phi)}{d\phi} &=& 0, \\
\ddot{\sigma}+3H\dot{\sigma}+\frac{dV(\sigma)}{d\sigma}
 &=& 0, \label{6},
\end{eqnarray}
where, H is the Hubble parameter, $H\equiv\dot{a}/a$. The first
Friedmann equation is given by,
 \begin{equation}
 H^{2}=\frac{2}{3}\rho.
 \end{equation}
 Substitute $\rho$ into above equation we obtain
\begin{equation}\label{fr1}
H^{2}=\frac{1}{3}\left[-\dot{\phi}^{2} +\dot{\sigma}^{2}+2V(\phi,
\sigma)\right],
\end{equation}
\begin{equation}\label{se}
\dot{H}=(\dot{\phi}^{2}-\dot{\sigma^{2}}).
\end{equation}
\section{First-order formalism}
 Now we introduce a first-order formalism, which shows how to relate the
potential that specifies the scalar field model to Hubble's
parameter. In order to obtain the first-order equation, we use
\cite{bez}
\begin{equation}\label{seq}
H=W, \hspace{1cm} \dot{\phi}=+W_{\phi}, \hspace{1cm}
\dot{\sigma}=-W_{\sigma}.
\end{equation}
From Eqs.(\ref{fr1}), (\ref{seq}) the explicit form of the potential
is
\begin{equation}\label{poteq}
V(\phi,\sigma)=\frac{3}{2}W^2+\frac{1}{2}(W_{\phi}^{2}-W_{\sigma}^{2}),
\end{equation}
where the super - potential $W$ is a well behaved function in the
space of scalar fields $\phi(x, t)$  and  $\sigma(x, t)$ $\in Maps
(R^{1, 1}, R^2).$ Here we assume that the superpotential be additive
as $W(\phi, \sigma) = W_{1}(\phi)+W_{2}(\sigma)$, so we have,
\begin{equation}\label{phi}
W_{\phi \sigma}=W_{\sigma\phi}=0.
\end{equation}
The equations (\ref{seq}) are first-order differential equations,
and they consistent with the set of equations (\ref{5}, \ref{6}) and
(\ref{fr1}, \ref{se}) for the potential (\ref{poteq}). The
constraint (\ref{phi}) guide us to consider a super-potential with
the following form \cite{{gom},{sade}}:
\begin{equation}\label{super1}
 W(\phi, \sigma)=\sinh K_{\phi}\phi+\sin K_{\sigma}\sigma
\end{equation}
where $K_{\phi}$ and $K_{\sigma}$ are constant. Using
Eq.(\ref{poteq}), we get the following potential,

\begin{equation}\label{poteq1}
V(\phi,\sigma)=\frac{3}{2}[\sin (K_{\sigma}\sigma)+ \sinh
(K_{\phi}\phi)]^{2}+\frac{1}{2}(
K_{\phi}^{2}\cosh^{2}(K_{\phi}\phi)-
K_{\sigma}^{2}cos^{2}(K_{\sigma}\sigma)).
\end{equation}
Now to obtain $\phi$ and $\sigma$ in term of $t$, we use equation
(\ref{seq})
\begin{equation}\label{eq1}
\frac{d\phi}{dt} = W_{\phi} = K_{\phi}\cosh (K_{\phi}\phi),
\end{equation}
\begin{equation}\label{eq2}
\frac{d\sigma}{dt} = -W_{\sigma} = -K_{\sigma}\cos
(K_{\sigma}\sigma).
\end{equation}
Now we are going to obtain $\phi(t)$ and $\sigma(t)$,
\begin{equation}\label{fi}
\phi(t) = \frac{1}{K_\phi}\ln\left[\tan(\frac{{K_\phi}^2
t}{2})\right],
\end{equation}
and
\begin{equation}\label{si}
\sigma(t) = \frac{1}{K_\sigma}\sin^{-1}\left[\tanh({-K_\sigma}^2
t)\right].
\end{equation}

 Using the above equations, we can rewrite the super-potential (\ref{super1}) as function of $t$ ,
\begin{equation}
W(t)=H(t) = \sinh \left[\ln(\tan\frac{{K_\phi}^2
t}{2})\right]+\tanh(-K_{\sigma}^2 t).
\end{equation}

As we know $W(t)=H(t)$, then we can obtain the energy density as a
fact in of time,
\begin{equation}
\rho(t) = \frac{3}{2}H^{2}(t) = \frac{3}{2}W^2(t) =\frac{3}{2}
(\sinh\left[\ln(\tan\frac{{K_\phi}^2
t}{2})\right]+\tanh(-K_{\sigma}^2 t))^{2}.
\end{equation}
Also from Eqs.(4, \ref{fi}, \ref{si}) one can obtain the pressure as
function of time :
\begin{eqnarray}
P(t) &=&
-\frac{3}{2}W^{2}(t)+\left[W^{2}_{\sigma}(t)-W^{2}_{\phi}(t)\right]=
K_{\sigma}^{2}\cos^{2}(K_\sigma \sigma)-K_{\phi}^{2}\cosh^{2}(K_\phi
\phi)\nonumber\\&-&\frac{3}{2}\left[\sinh^{2}(K_{\phi}\phi)+\sin^{2}(K_{\sigma}\sigma)+
2\sinh(K_{\phi}\phi)\sinh(K_{\sigma}\sigma)\right].
\end{eqnarray}
Now we are going to write the equation of state as follow,
\begin{equation}
\omega
=\frac{P(t)}{\rho(t)}=-1+\frac{2(W_{\sigma}^2-W_{\phi}^2)}{3W^{2}}.
\end{equation}
and the acceleration parameter $q(t)$  given by,
\begin{equation}
q(t)= 1+\frac{\dot H}{H^2}=1+\frac{W_{\phi}^2-W_{\sigma}^2}{W^{2}}
\end{equation}
In Figure 1 we plot $H(t)$, $\omega(t)$ and $q(t)$ for some choice
of parameters.\\
Substituting Eqs.(\ref{fi}), and (\ref{si}) respectively in
Eqs.(\ref{eq1}), (\ref{eq2}) one can obtain
\begin{equation}\label {eq3}
W_{\phi}=\frac{K_{\phi}(tan^{2}(\frac{K_{\phi}^{2}t}{2})+1)}{2
tan(\frac{K_{\phi}^{2}t}{2})}
\end{equation}
\begin{equation}\label {eq4}
W_{\sigma}=K_{\sigma}\sqrt{1-tanh^{2}(K_{\sigma}^{2}t)}
\end{equation}
If $W_{\phi}^2<W_{\sigma}^2$, then $\omega >-1$, in this case we are
in quintessence phase, in the other hand if
$W_{\phi}^2>W_{\sigma}^2$, then $\omega <-1$ in this case the
universe is in the phantom phase.\\
Therefore we have extended the first-order formalism introduced in
\cite{{bez},{bez2},{2}} to describe the FRW cosmology, driven by a
couple of  scalar fields $\sigma$, $\phi$ with  standard  dynamics
for flat spatial geometry. The present method may be used to
investigate several interesting cases, in particular the case in
which the cosmic evolution occurs in closed or open geometry, for
phantom or quintom models.

\begin{tabular*}{5cm}{cc}
\hspace{0.25cm}\includegraphics[scale=0.25]{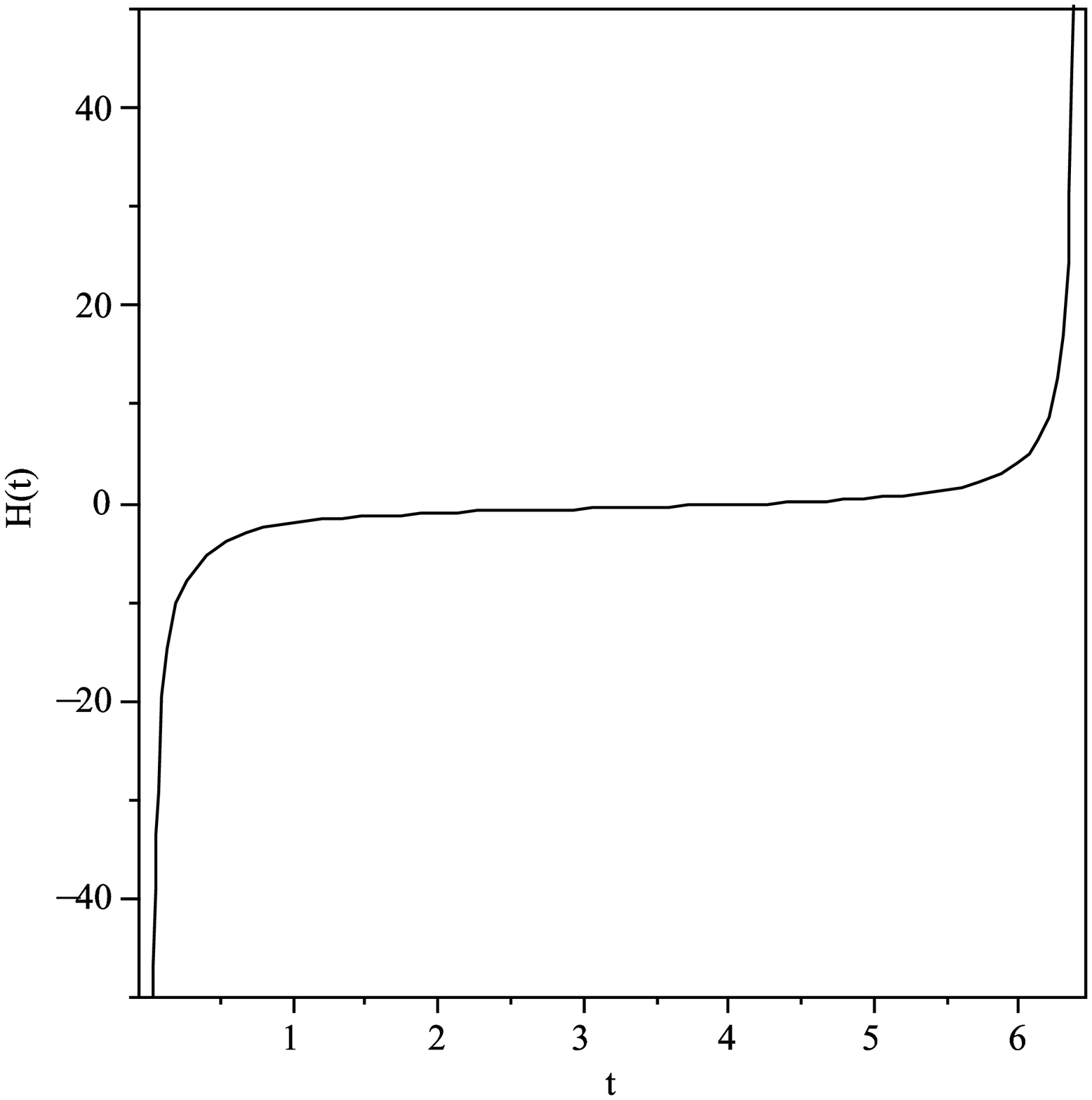}\hspace{0.5cm}\includegraphics[scale=0.25]
{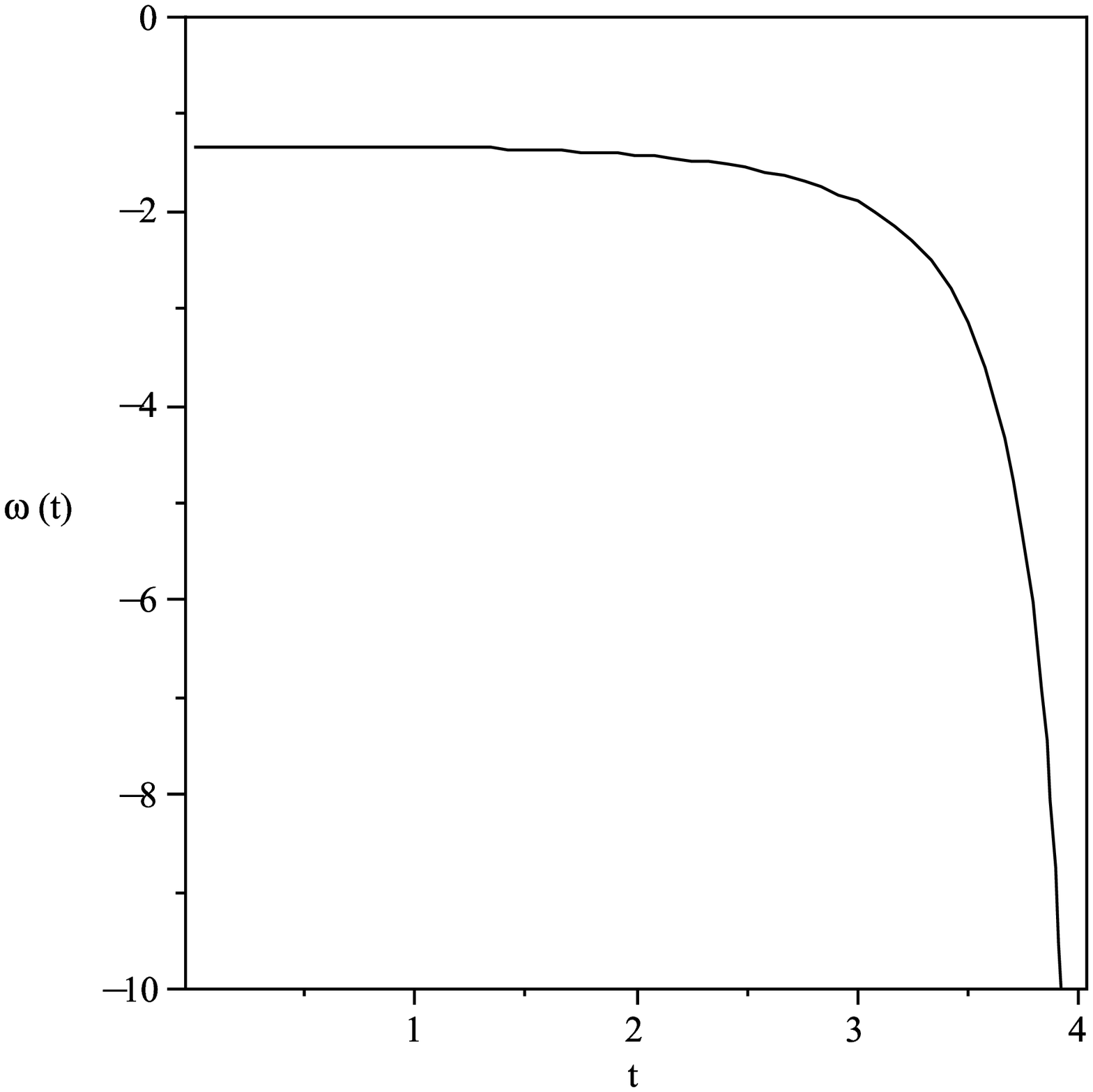}\hspace{0.5cm}
\includegraphics[scale=0.25]{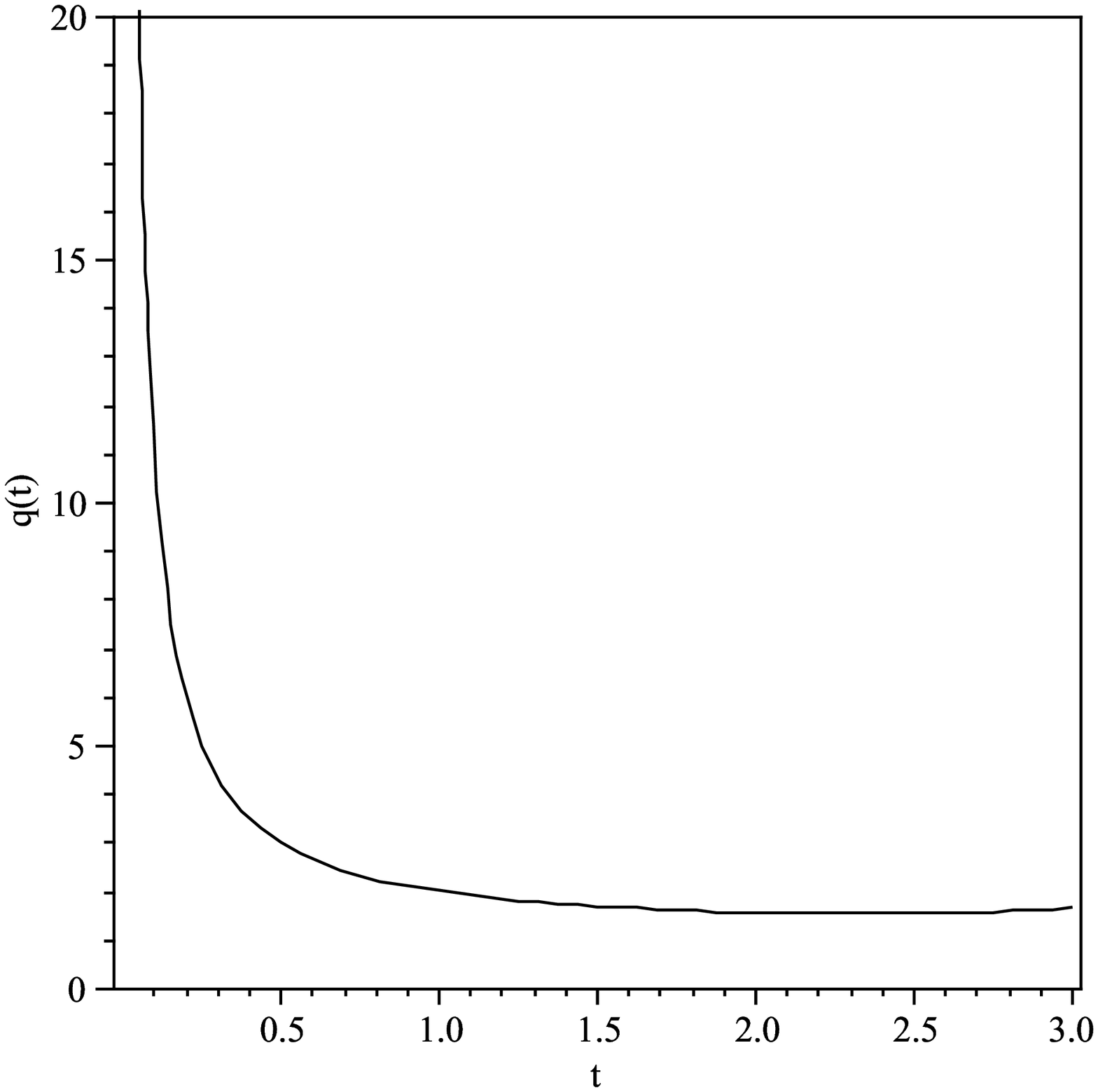}\\
\hspace{-1.5cm}\textbf{Figure 1:} \,The Graphs of $H(t)$,
$\omega(t)$
and $q(t)$ draw in term of $t$ for $K_\phi=0.7$ and $K_\sigma=0.4$. \\
\end{tabular*}\\\\\\

\section{conclusion}
The idea to consider the Hubble parameter as a function of scalar
fields and to transform Eqs.(6,7,9,10) into Eqs.(11,12) has been
used in the Hamilton–Jacobi formulation of the Friedmann equations
and does not connect with supersymmetric and supergravity theories
\cite{1}. Also the idea to apply system (11,12) instead of the
original equations of motion and to seek in such a way exact special
solutions is actively used in two-dimensional fields models
\cite{2}. In the present work we have shown how to write a
first-order formalism to FRW cosmology and to the quintom model of
dark energy with two scalar fields. The crucial ingredient was the
introduction of a new function, $W=W(\phi, \sigma)$ from which we
could express Hubble's parameter in the form $H(t)=W[\phi(t),
\sigma(t)]$. Also the energy density and pressure can be obtained by
$H(t)=W[\phi(t), \sigma(t)]$.  Finally by using the energy density
and pressure we obtained the equation of state for the quintom
model. The condition for the accelerated expansion is obtained by
equation of state. The deformation procedure for two scalar field in
quintom model of dark energy may be intersecting for future work.

 \end{document}